\documentclass[twocolumn,tighten]{aastex61}

\usepackage[sort&compress]{natbib}
\bibpunct{(}{)}{;}{a}{}{,}
\usepackage{color}
\definecolor{darkgreen}{RGB}{0,142,128}

\usepackage{hyperref}
\hypersetup{colorlinks,citecolor=blue,linkcolor=blue}

\usepackage{amsmath}
\usepackage{amsthm}
\usepackage{amsfonts}
\usepackage{url}
\usepackage{subfigure}
\usepackage{multirow}

\newcommand{\modd}[1]{{#1}}
\newcommand{\moddd}[1]{{#1}}
\newcommand{\reff}[1]{{#1}}
\newcommand{\refff}[1]{{#1}}
\newcommand{\reffff}[1]{{#1}}

\begin{document}

\title{Chasing star-planet magnetic interactions: \\ the case of Kepler-78}
\shorttitle{Chasing star-planet magnetic interactions: the case of Kepler-78}

\author{A. Strugarek}
\affil{AIM, CEA, CNRS, Universit\'e Paris-Saclay, Universit\'e Paris Diderot, Sorbonne Paris Cit\'e, F-91191 Gif-sur-Yvette, France}
\email{antoine.strugarek@cea.fr}
\author{A. S. Brun}
\affil{AIM, CEA, CNRS, Universit\'e Paris-Saclay, Universit\'e Paris Diderot, Sorbonne Paris Cit\'e, F-91191 Gif-sur-Yvette, France}
\author{J.-F. Donati}
\affil{Universit\'e de Toulouse / CNRS-INSU, IRAP / UMR 5277, F-31400 Toulouse, France}
\author{C. Moutou}
\affil{Canada-France-Hawaii Telescope Corporation, CNRS, 65-1238, Mamalahoa Hwy, Kamuela, HI 96743, USA}
\author{V. R{\'e}ville}
\affil{UCLA Earth, Planetary and Space Sciences, 595 Charles E. Young Drive East, 90095 Los Angeles, CA, vreville@epss.ucla.edu}

\shortauthors{A. Strugarek \textit{et al.}}
\begin{abstract}
Observational evidence of star-planet magnetic interactions (SPMI) in compact exo-systems have been looked for in the past decades. Indeed planets in close-in orbit can be magnetically connected to their host star, and channel Alfv\'en waves carrying large amounts of energy towards the central star. The strength and temporal modulation of SPMIs are primarily set by the magnetic topology of the host star and the orbital characteristics of the planet. As a result, SPMI signals can be modulated over the rotational period of the star, the orbital period of the planet, or a complex combination of the two. The detection of SPMI thus have to rely on multiple-epochs and multiple-wavelengths observational campaigns. We present a new method to characterize SPMIs and apply it to Kepler-78, a late G star with a super-Earth on an 8.5 hours orbit. We model the corona of Kepler-78 using the large-scale magnetic topology of the star observed with Zeeman-Doppler-Imaging. We show that the closeness of Kepler-78b allows the interaction to channel energy flux densities up to a few kW m$^{-2}$ towards the central star. \refff{We show that this flux is large enough to be detectable in classical activity tracers such as H$\alpha$. It is nonetheless too weak to explain the modulation observed by \citet{Moutou2016}. We furthermore demonstrate how to predict the temporal modulation of SPMI signals in observed systems such as Kepler-78.} The methodology presented here thus paves the road towards denser, specific observational campaigns that would allow a proper identification of SPMIs in compact star-planet systems.
\end{abstract}



\section{Introduction}
\label{sec:introduction}

Planets on short-period orbit are expected to interact strongly with their host star \citep{Cuntz2000} due to the strong irradiation they receive \citep{Yelle2008,Trammell2014,Matsakos2015,Bourrier2016,Daley-Yates2018}, the strong tidal forces they experience \citep{Mathis2017b,Bolmont2017a,Strugarek2017c}, and the strong inter-planetary magnetic field they orbit in \citep[see][and references therein]{Lanza2010,Cohen2011,Strugarek2016c,Strugarek2017b}. Such close-in planets will generally orbit in the sub-alfv\'enic region of the accelerating wind of their star. 
In that case they can magnetically back-react on their host, which is often referred to as star-planet magnetic interaction (SPMI). \modd{In exoplanetary systems the SPMIs are expected to channel magnetic energy carried by Alfv\'en waves excited by the interaction, and travelling from the planet to the star. The superposition of these waves form the so-called Alfv\'en wings. Alfv\'en wings are a very robust feature of SPMIs as they are triggered whether or not the planet's atmosphere escapes, and whether or not the planet posseses a magnetosphere \citep{Neubauer1998}. SPMIs were shown to be able to channel magnetic powers up to 10$^{19}$-10$^{20}$ W in the most favorable star-planet systems \citep{Saur2013,Strugarek2016c}. These large energy fluxes can in principle leave observable traces in stellar atmospheres, through \textit{e.g.} local heating \citep{Ip2004}, flare trigger \citep{Lanza2018,Fischer2019}, or particle acceleration \citep{Hess2011} as in the case of Io-Jupiter magnetic interaction \citep{Clarke2002a}.}

\modd{Numerous attempts to observe SPMIs signals have been carried out in the past decade, for specific star-planet systems \citep[e.g.][and references therein]{Fares2010,Figueira2016,Zarka2017,Fischer2019} as well as for ensembles of star-planet systems \citep{Poppenhaeger2011,Miller2015}. None of these studies were able to unambiguously detect SPMIs. Recently, \citet{Cauley2018} reported a first detection in the CaII K band of HD 189733, where an anomalous flux correlated with the orbital period of the planet was observed. This signal was not detected in previous observational campaigns of HD 189733 \citep{Fares2010}.}

\modd{At first order, SPMIs depend only on the magnetic properties of the host star as was shown unambiguously by \citet{Strugarek2015}. This dependancy naturally leads to on/off SPMIs \citep{Shkolnik2008} as the large-scale stellar magnetic field changes over decadal timescales. More precisely} the magnetic topology of the corona of the star determines where the foot-point of the interaction lies on the star, and thus where any tracer of the interaction would actually be localized. It thus appears clearly that any attempt to characterize observable effects of SPMIs on the stellar flux must take into account the magnetic topology of the central star. 

Fortunately, Zeeman-Doppler Imaging \citep[ZDI, see][]{Donati2009} has become in the past decades an extremely valuable tool to characterize the large-scale magnetic topology of bright stars thanks to spectro-polarimetric observations. Leveraging these constraints for given star-planet systems, it is now possible to predict both the energetic and the temporal variability of SPMIs. 

In this paper we present a \moddd{new method to characterize and predict star-planet magnetic interactions. We apply the method to SPMIs for Kepler-78,}  an ultra-short period star-planet system with an orbital period of 0.36 days \citep{Sanchis-Ojeda2013}. We demonstrate that SPMI channels enough power to be detectable in classical magnetic activity tracers such as the CaII H\&K or H$\alpha$ bands. We carry out a detailed modelling of the magnetized wind of Kepler-78 to assess the temporal variability of the SPMI signal. 
\moddd{Our model predicts a specific temporal modulation over the orbital and rotational timescales of the system, and we show that SPMIs \refff{could be detectable} in the CaII H\&K or H$\alpha$ bands of Kepler-78. We show that the overall method can provide a powerful and robust guide to design dedicated observational campaigns to detect SPMIs during the most favorable rotational and orbital phases of  star-planet systems.}

The paper is organized as follows. We first introduce in \S\ref{sec:kepler-78-system} the characteristics of the Kepler-78 system, along with the activity tracers observed by \citet{Moutou2016}. We then present in \S\ref{sec:simul-magn-kepl} simulations results of the corona of Kepler-78 based on the ZDI maps of the star as of July 2015. In \S\ref{sec:prop-stell-wind} we use the simulations results to predict the properties of the star-planet magnetic interaction along the orbit of Kepler-78b. We further estimate in \S\ref{sec:energ-star-plan} the energetic and temporal variability of the magnetic interaction. We finally discuss in \S\ref{sec:discussions} the implications of this study for future prospects in characterizing star-planet magnetic interaction is compact star-planet systems, and summarize our results and conclude in \S\ref{sec:conclusions-1}.

\section{The Kepler-78 system}
\label{sec:kepler-78-system}

Kepler-78 and its close-in planet Kepler-78b form one of the ultra-short-period orbital systems that has been discovered in the past decade \citep{Winn2018}. It consists of an Earth-like planet around a late G--early K star with an orbital period of 0.36 days. This planet was first reported by \citet{Sanchis-Ojeda2013}, and we list in Table \ref{ta:k78prop} the characteristics of the host star and its planets found in the literature. The very short orbit of Kepler-78b makes it a fantastic testbed for star-planet tidal and magnetic interactions \citep{Strugarek2017b,Mathis2017b}. 

\begin{deluxetable}{lr}
  \tablecaption{Properties of the Kepler-78 system\label{ta:k78prop}}
  \tablecomments{The parameters of the Kepler-78 system were published in
	\citet{Sanchis-Ojeda2013}. The error bars correspond to the 63.8\%
    confidence limits \citep[see][]{Sanchis-Ojeda2013}. $^\dagger$ This
    mass that was deduced
    from HARPS-N data by \citet{Pepe2013}. \citet{Howard2013}
    and \citet{Sanchis-Ojeda2013} give much larger error bars on the
    mass, with a maximum value of 8 $M_\oplus$. $^{\dagger\dagger}$ The semi-major axis of Kepler-78b was reported to be $3\,R_\star$ in \citet{Sanchis-Ojeda2013}, with significant error bars. Applying Kepler's law for a circular orbit, we find that $R_{\rm orb} \sim 2.66\, R_\star$, based on the well constrained orbital period of Kepler-78b. We use the latter value in this work.}
  \tablecolumns{2}
  \tabletypesize{\scriptsize}
  \tablewidth{\linewidth}
  \tablehead{
    \colhead{Parameter} &
    \colhead{\hspace{1.cm}Value}\hspace{1.cm} 
  }
  \startdata 
  $T_{\rm eff}$ $[K]$   & 5089 $\pm$ 50      \\ 
  $M_\star$ $[M_\odot]$ & 0.81 $\pm$ 0.08    \\ 
  $R_\star$ $[R_\odot]$ & 0.74 +0.1,-0.08     \\ 
  $P_{\rm rot}$ [days]  & 12.5               \\
  \hline
  $R_p$ [$R_\oplus$]       & 1.16 +0.19,-0.14 \\
  $M_p^{\dagger}$ [$M_\oplus$]       & 1.86 $\pm$ 0.25  \\
  $P_{\rm orb}$ [days]      & 0.36 \\
  Semi-major axis [$R_\star$] & 2.66$^{\dagger\dagger}$\\
  \enddata
\end{deluxetable}

After its discovery, follow-up studies of Kepler-78 were carried out by \citet{Moutou2016} to characterize its magnetic properties. Using spectropolarimetric observations with Canada-France-Hawaii Telescope (CFHT)/ESPaDOnS, they measured the circular polarization in the line profile of the star and mapped the photospheric magnetic field using Zeeman-Doppler Imaging (ZDI). They further gathered spectroscopic proxies of the chromospheric activity of Kepler-78, which revealed a \moddd{peculiar} modulation with the rotational phase of the star. \citet{Moutou2016} observed a significant change in the H and K bands of CaII, in H$\alpha$, H$\beta$, and in the infra-red triplet of CaII. In this work we will focus on H$\alpha$ for the sake of simplicity, the other tracers giving qualitatively similar results \reff{(see Appendix \,\ref{sec:activ-trac-kepl})}. 

The excess and dearth of H$\alpha$ was reported in normalized units in Table 1 of \citet{Moutou2016} for a series of 13 observations between July 2015 and August 2015. We recast here these variations using the following procedure. First, the smallest value of \moddd{the H$\alpha$ line is chosen as the zero-level absorption}. We will assume here that the continuum spectrum of Kepler-78 is well described by a black-body with $T_{\rm eff}= 5089$ K (see Table \ref{ta:k78prop}). \refff{In the H$\alpha$ band the black-body approximation to the continuum flux does not introduce an error larger than about 10\% when compared to more realistic spectra \citep[e.g.][]{Husser2013}.
\reffff{Second, the black-body radiance at $\lambda_{H\alpha}=6563${\AA} is integrated over the width of the residual line as in \citet{Moutou2016}.} Third, we integrate the result over all \modd{emission angles} and obtain the observed flux. This procedure allows to quantify a flux hypothetically distributed over the whole stellar disk. If the source is localized over a given area $A$ on the stellar disk, this observed flux must be scaled by the relative area $A/(\pi R_\star^2)$. This signal could in principle originate from stellar activity, SPMIs, or a combination of both.} We explore in this work the possibility of a planetary origin.

\begin{figure}[htb]
  \centering
  \includegraphics[width=\linewidth]{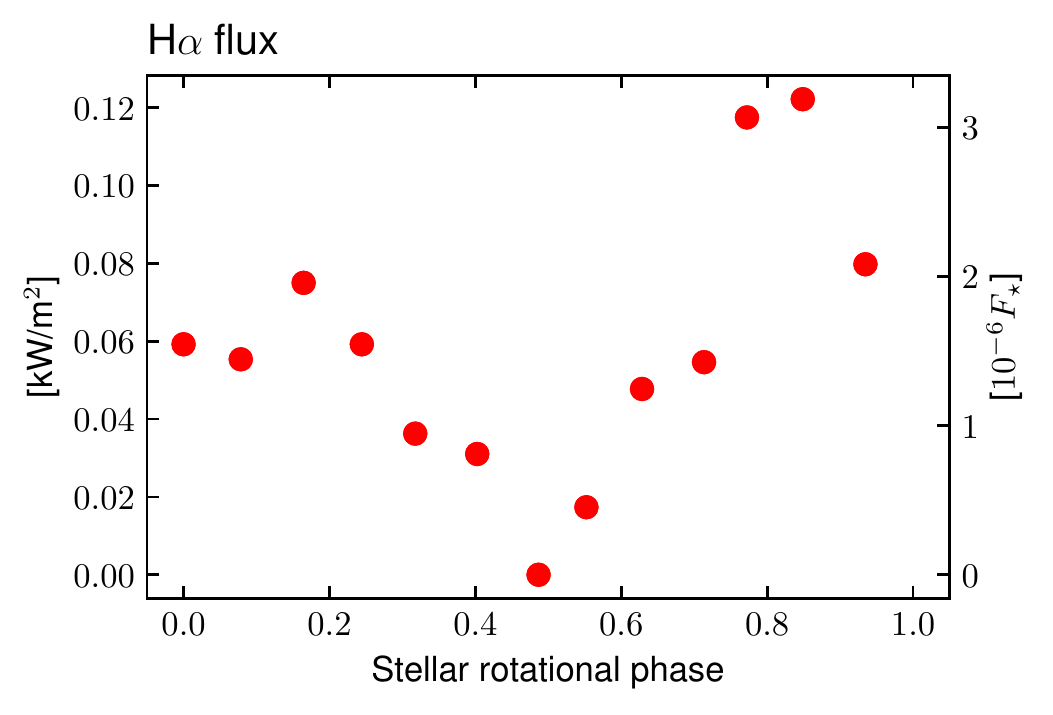}
  \caption{Flux observed in H$\alpha$ reported by \citet{Moutou2016}, as a function of the stellar rotational phase of the star. Observations taken on different rotation cycles have been phase-folded on one normalized rotational period set to 12.5 days (see Table \ref{ta:k78prop}). A black-body at $T_{\rm eff}=5089$ K has been assumed for Kepler-78. The flux is obtained by removing the quieter signal during the observation campaign (see text). It is shown in units of kW m$^{-2}$ (left vertical axis), and relative to the stellar bolometric flux $F_\star$ (right vertical axis). \refff{The flux was assumed here to be isotropically distributed over the stellar disk.}} 
  \label{fig:equivlums}
\end{figure}

We show the observed flux in Fig. \ref{fig:equivlums}, where we phase-folded all the observations published in \citet{Moutou2016}. \modd{The phase-folding must  be taken with caution as we did not correct the flux for any intrinsic variability of the star nor from the latitudinal differential rotation.} We scaled the H$\alpha$ flux in absolute units (left axis) and in units of the stellar bolometric flux $F_{\star}=L_\star/(4\pi R_\star^2)$ (right axis, we estimated $L_\star \simeq 0.33 L_\odot$ assuming the radius and effective temperature reported in Table \ref{ta:k78prop}). The flux of Kepler-78 in H$\alpha$ reaches a few hundreds of W m$^{-2}$. The power needed to explain this flux depends on the size of the area at the source of this signal. In the remainder of this paper we will characterize the strength and size of a hot-spot magnetically induced by Kepler-78b to assess whether this signal can originate from SPMI or not. 

\section{Models of the wind of Kepler-78}
\label{sec:simul-magn-kepl}

We model the wind of Kepler-78 using the same approach than
\citet{Reville2016a}, based on the Zeeman-Doppler Imaging (ZDI)
map of Kepler-78 observed in July 2015 \citep[][]{Moutou2016}. \reff{The radial component of the surface magnetic field of Kepler-78 reaches typical values of 3 mT (30 G, see Fig. \ref{fig:FootPoints}).} Kepler-78 rotates a bit more than twice as fast as the Sun, with a Rossby number of approximately 0.1 \citep[the convective turnover time was taken from the parametrization given in][]{Gunn1998,Cranmer2011}. \moddd{As a result Kepler-78 is close to saturation \citep[e.g.][]{Matt2015,Johnstone2015}, and its coronal properties can be estimated with} empirical formulae using power-law exponents derived by \citet{Holzwarth2007} \citep[see also][]{Reville2016a}, \textit{i.e.}
\begin{equation}
  \label{eq:rho_and_T}
  T_c = T_\odot\left(\frac{\Omega}{\Omega_\odot}\right)^{0.1}\, , \,\,\,
  n_c = n_\odot \left(\frac{\Omega}{\Omega_\odot}\right)^{0.6}\, ,
\end{equation}
where $T_\odot = 1.5\times 10^6 $K, $n_\odot = 10^8 $ cm$^{-3}$ and
$\Omega_\odot = 2.6 \times 10^{-6}$ s$^{-1}$. \modd{We obtain $T_c=1.63 \times 10^6$ K and $n_c = 1.6 \times 10^8$ cm$^{-3}$ for the corona of Kepler-78.} 
The sensitivity of our results with respect to this choice of coronal parameters is discussed in \S\ref{sec:discussions}.

We first model the wind of Kepler-78 using the starAML
package published in \citet{Reville2015} and accessible upon request. This package solves the
pressure balance of a magnetized 1D Parker-like wind \citep{Parker1958a,Weber1967,Sakurai1985} to calculate an
optimal source-surface, which is in turn used to estimate the
magnetic open flux deduced from the observed ZDI maps of
the magnetic field at the surface of the star. 

The optimal source-surface for Kepler-78 is found to be $R_{ss} = 6.9 R_\star$. The three-dimensional structure of the magnetic field in the low corona of the star can then be estimated using a potential-field source surface (PFSS) extrapolation method \citep{Schrijver2003a}. \reff{The chosen source-surface radius ensures that the PFSS extrapolation of the magnetic field matches as closely as possible the magnetic topology of the corona obtained with an MHD model \citep{Reville2015}.} 

We refer to this coupled model as starAML+PFSS, which is conveniently very inexpensive numerically. \reff{It is worth noting that this model is multi-D: the starAML Parker-like wind model is 1D but the magnetic topology from the PFSS extrapolation is 3D \citep[see][for the details
concerning this procedure]{Reville2015}.} Similar multi-D approaches have been recently explored to model in near real-time the solar wind \citep{Pinto2017}. 
The resulting global properties
of the wind of Kepler-78, based on the ZDI map from July 2015, are given in the first column of table \ref{ta:k78wind}. We find that Kepler-78 has mass loss $\dot{M}$ and angular momentum loss $\dot{J}$ similar to the Sun (recall that $\dot{M}_\odot \simeq 2-3 \times 10^{-14} M_\odot/yr$, see \citealt{McComas2008}), and an averaged Alfv\'en radius $\left\langle r_A \right\rangle = \sqrt{\frac{\dot{J}}{\Omega_\star \dot{M}}}$ also similar to the Sun \citep[e.g.][]{Reville2016a}. 

The complex topology of the magnetic field of Kepler-78
warrants a more detailed investigation of the structure of its wind,
especially in the context of SPMIs. We
have hence computed a 3D wind solution of Kepler-78 using the PLUTO code \citep{Mignone2007} under the magneto-hydrodynamical (MHD) formalism \citep{Strugarek2014a,Reville2016a}, based
on the same ZDI map of Kepler-78 \citep{Moutou2016}. The averaged properties of the 3D simulated wind are reported in the last column of Table \ref{ta:k78wind}. The averaged 3D model and starAML+PFSS agree very well (see \S\ref{sec:wind-prop-along} for their modulation along the planetary orbit), which shows that simplified multi-D approaches already provide a very useful first estimate of the averaged properties of stellar winds. 

\begin{figure}[htb]
  \centering
  \includegraphics[width=\linewidth]{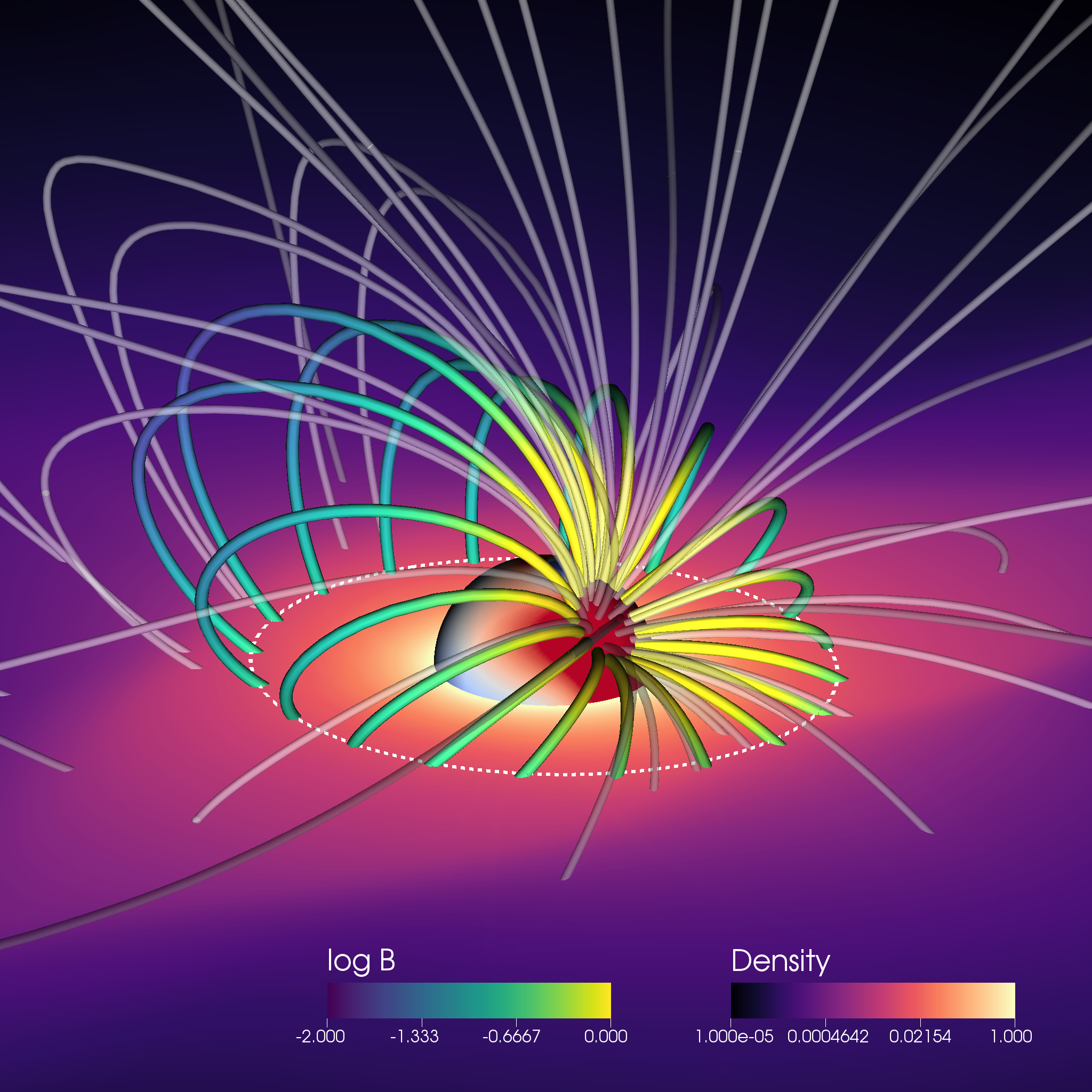}
  \caption{Three-dimensional visualization of the environment of Kepler-78 \moddd{at rotational phase $\phi_{\rm rot}\sim 0.3$}. 
    The radial magnetic field at the surface of the star is represented by the blue/red (negative/positive) sphere in the center. The density in the wind is shown on the orbital plane in logarithmic scale with a black-purple-white color table. The magnetic field lines of Kepler-78 are represented by the transparent gray tubes. The orbit of Kepler-78b is shown by the dashed white circle, and the magnetic field lines connecting the orbit of the planet to the stellar surface are highlighted by the colored field lines.}
  \label{fig:3DWind}
\end{figure}

We illustrate the 3D wind steady-state solution of Kepler-78 in Figure \ref{fig:3DWind}, \reffff{seen at rotational phase $\phi_{\rm rot}\sim 0.3$ (the rotational phase is defined as in \citealt{Moutou2016}, with $\phi_{\rm rot}=0$ corresponding to the observation made on July 23rd, 2015)}. The density in the wind is shown by the purple colormap on the orbital plane, and the orbit of Kepler-78b is labeled by the white dashed line. The stellar wind magnetic field lines are shown as grey and colored tubes. The colored field lines show the magnetic connectivity between the planet and the star. 

We now study in details the properties of SPMI along the orbit of Kepler-78b using the 3D model of the wind of Kepler-78.

\begin{deluxetable}{lll}
  \tablecaption{Simulations of the wind of Kepler-78.\label{ta:k78wind}}
  \tablecomments{Both 1D and 3D leverage the magnetic topology of Kepler-78, as observed in July 2015. The calculation of the mass loss ($\dot{M}$), angular momentum loss ($\dot{J}$), and averaged Alfv\'en radius $\left\langle r_A\right\rangle$ are detailed in \citet{Reville2016a}.}
  \tablecolumns{3}
  \tablewidth{\linewidth}
  \tabletypesize{\scriptsize}
  \tablehead{
    \colhead{Parameter} &
    \colhead{starAML+PFSS} &
    \colhead{3D wind} 
  }
  \startdata 
  $\dot{M}$ $[M_\star/yr]$ & 3.56 $\times$ 10$^{-14}$ & 2.13 $\times$
  10$^{-14}$ \\
  $\dot{J}$ $[R_\star^2 M_\star/yr^2]$ & 4.54 $\times$ 10$^{-10}$ & 3.73 $\times$ 10$^{-10}$\\
  $\left\langle r_A \right\rangle$ $[R_\star]$ & 11.0 & 9.81 \\
  \enddata
\end{deluxetable}

\section{Star-planet magnetic interaction along the planetary orbit}
\label{sec:prop-stell-wind}

\subsection{Wind properties along the orbit}
\label{sec:wind-prop-along}

\begin{figure}[htb]
  \centering
  \includegraphics[width=\linewidth]{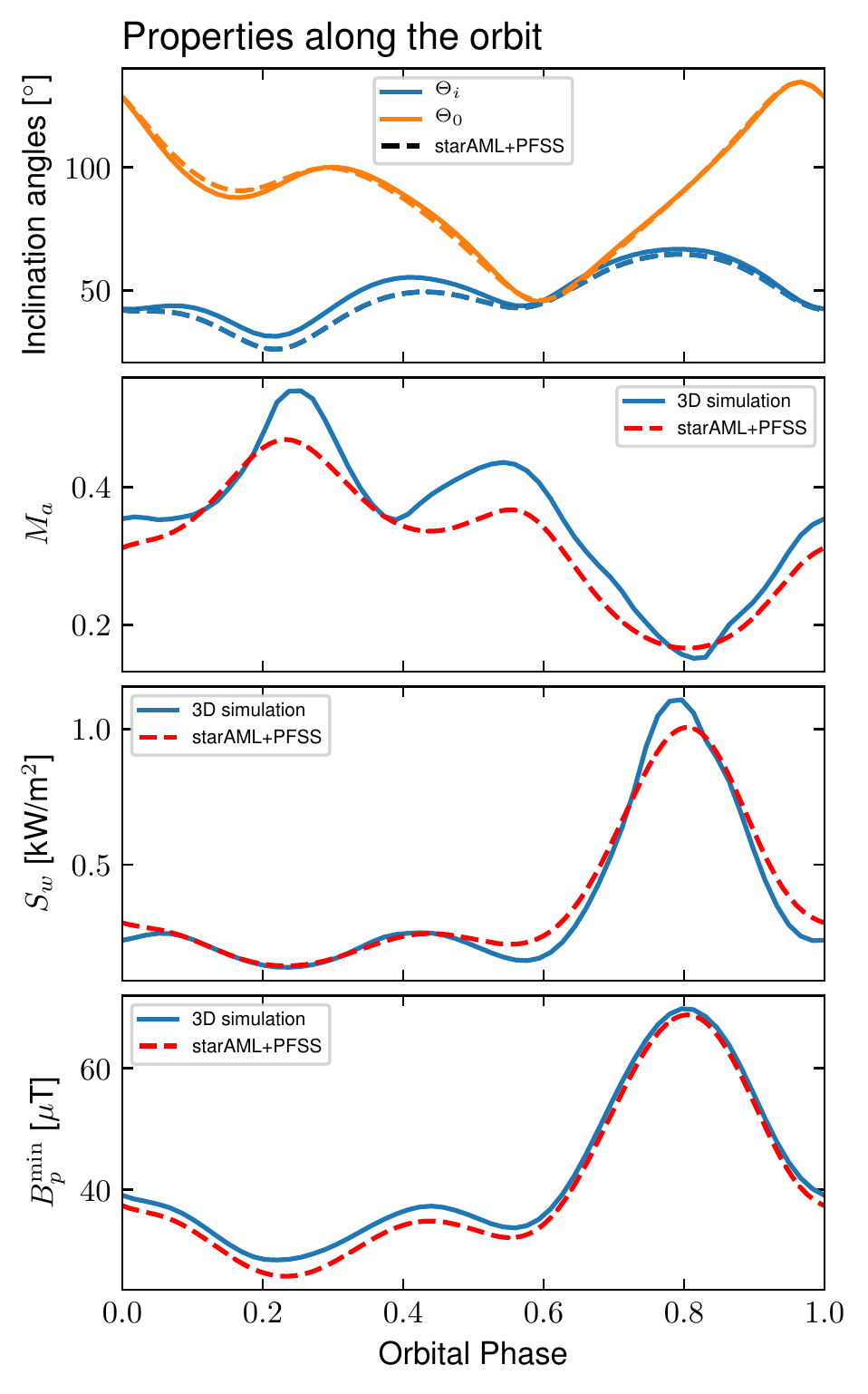}
  \caption{Properties of the stellar wind along the orbit of Kepler-78b. The inclination angles of the stellar wind magnetic field with respect to the orbital plane (blue) and the orbital motion (orange) are shown in the first panel. The alfv\'enic Mach number of the SPMI is shown in the second panel, and the Poynting vector in the third panel. Finally, the minimum planetary surface field required to sustain a magnetosphere is shown in the last panel. In all the panels, the blue line correspond to the 3D stellar wind model of Kepler-78b. The values for the simplified stellar wind starAML+PFSS model are indicated by the dashed lines.}
  \label{fig:PropAlongOrbit}
\end{figure}

The planet Kepler-78b orbits in an inhomogeneous medium due to the complex magnetic topology of the magnetic field of Kepler-78. The
magnetic interaction of the planet with its host star depends on the
plasma properties of the wind along its orbit. \modd{The Alfv\'en waves excited by the interaction form what is called the Alfv\'en wings of the magnetic interaction. These waves propagate in the $({\bf B}_w,{\bf v}_0)$ plane where ${\bf B}_w$ is the wind magnetic field, and ${\bf v}_0={\bf v}_w-{\bf v}_{\rm kep}$ is the differential motion between the rotating wind velocity  ${\bf v}_w$ and the keplerian orbital velocity ${\bf v}_{\rm kep}$. As a result,} the inclination of the wind magnetic field with respect to the orbital motion ($\Theta_0$) and with respect to \reff{the normal to the orbital plane ($\Theta_I$)} were shown to be important parameters determining the strength and shape of the magnetic interaction
\citep{Saur2013,Strugarek2016c,Strugarek2017e}. We show the inclination angles in the upper
panel of Figure \ref{fig:PropAlongOrbit}. $\Theta_0$ (in orange) varies from 45$^\circ$
to 134$^\circ$ along the orbit. The interaction strength is maximized when $\Theta_0=90^\circ$ and minimized when $\Theta_0=0^\circ$ or $\Theta_0=180$. As a result the amplitude of the SPMI for Kepler-78b varies strongly in efficiency along the orbit. 

We define the local Alfv\'enic Mach number as $M_a = v_0/v_A$, \reff{where $v_A=|{\bf v}_A| = |{\bf B}_w|/\sqrt{\mu_0\rho_w}$ is the local Alfv\'en speed that depends on the local wind magnetic field $B_w = |{\bf B}_w|$} and density $\rho_w$. The energy flux channeled by the
magnetic interaction depends on this local Alfv\'enic Mach number $M_a$ (shown in the second panel of Fig. \ref{fig:PropAlongOrbit}), and on the Poynting flux density 
\begin{equation}
S_w = v_0B_w^2 sin(\Theta_0)/\mu_0
    \label{eq:PoyW}
\end{equation}
shown in the third panel. \modd{The Poynting flux density quantifies the total input energy available for the magnetic interaction \citep{Saur2013}.} \modd{In both panels, the values obtained
with the starAML+PFSS model are indicated by the 
dashed line. This simplified (albeit calibrated) model gives a very good estimate
of all parameters, as seen in Table \ref{ta:k78orb} where we report
the interaction parameters averaged along the planetary orbit for the
two wind models.}

\modd{The largest difference is found for the alfv\'enic Mach number $M_a$ (second panel), as the MHD model allows the density profile to be more realistic and self-consistent in 3D (see Fig. \ref{fig:3DWind}). We further note that the starAML+PFSS model gives accurate results thanks to the optimized source-surface radius $R_{ss}=6.9 R_\star$ \citep{Reville2015} used for the potential field extrapolation in the corona of Kepler-78. When the canonical solar value $R_{ss} = 2.5 R_\odot$ (or $R_{ss}=2.5 R_\star$) is used the starAML+PFSS model gives significantly different results: Kepler-78b is then found to enter and leave the Alfv\'en surface along the orbit. This is not the case when the fully 3D solution is taken into account, as shown in the second panel of Fig. \ref{fig:PropAlongOrbit}. 

Finally, the calibrated starAML+PFSS model is very accurate here because Kepler-78b orbits very close to its host star \moddd{and in particular well within $R_{ss}$}. As the orbital period increases we expect this model to be less and less precise and a fully 3D model of the stellar atmosphere would then be necessary.}

\modd{We observe that $M_a$ and $S_w$ vary by more than a factor 2 along the orbit, which demonstrates the need of a 3D model to quantify properly SPMIs
in real, particular star-planet systems. We further note that $M_a$ remains below
one along the orbit, such that Kepler-78b remains in a sub-alfv\'enic
interaction regime at all times.}

It is already worth noting that the power density available to the magnetic interaction, $S_w$, reaches about 1 kW m$^{-2}$ at its maximum. This power density is larger than the signal seen in the H$\alpha$ flux of Kepler-78  (see Figure \ref{fig:equivlums}). With a simple order of magnitude estimate, we immediately see that magnetic interactions could be at the source of the observed signal.

\begin{deluxetable}{lll}
  \tablecaption{Averaged properties at the planetary orbit \label{ta:k78orb}}
  \tablecolumns{3}
  \tabletypesize{\scriptsize}
  \tablewidth{\linewidth}
  \tablehead{
    \colhead{Parameter} &
    \colhead{starAML+PFSS} &
    \colhead{3D wind} 
  }
  \startdata 
 $M_a$ & 0.32 & 0.35 \\
 $S_w$ $[kW/m^2]$ & 0.37 & 0.35 \\
 $B_p^{\rm min}$ $[\mu T]$ & 69 & 70 \\
  \enddata
\end{deluxetable}

\begin{figure*}[htb]
  \centering
  \includegraphics[width=\linewidth]{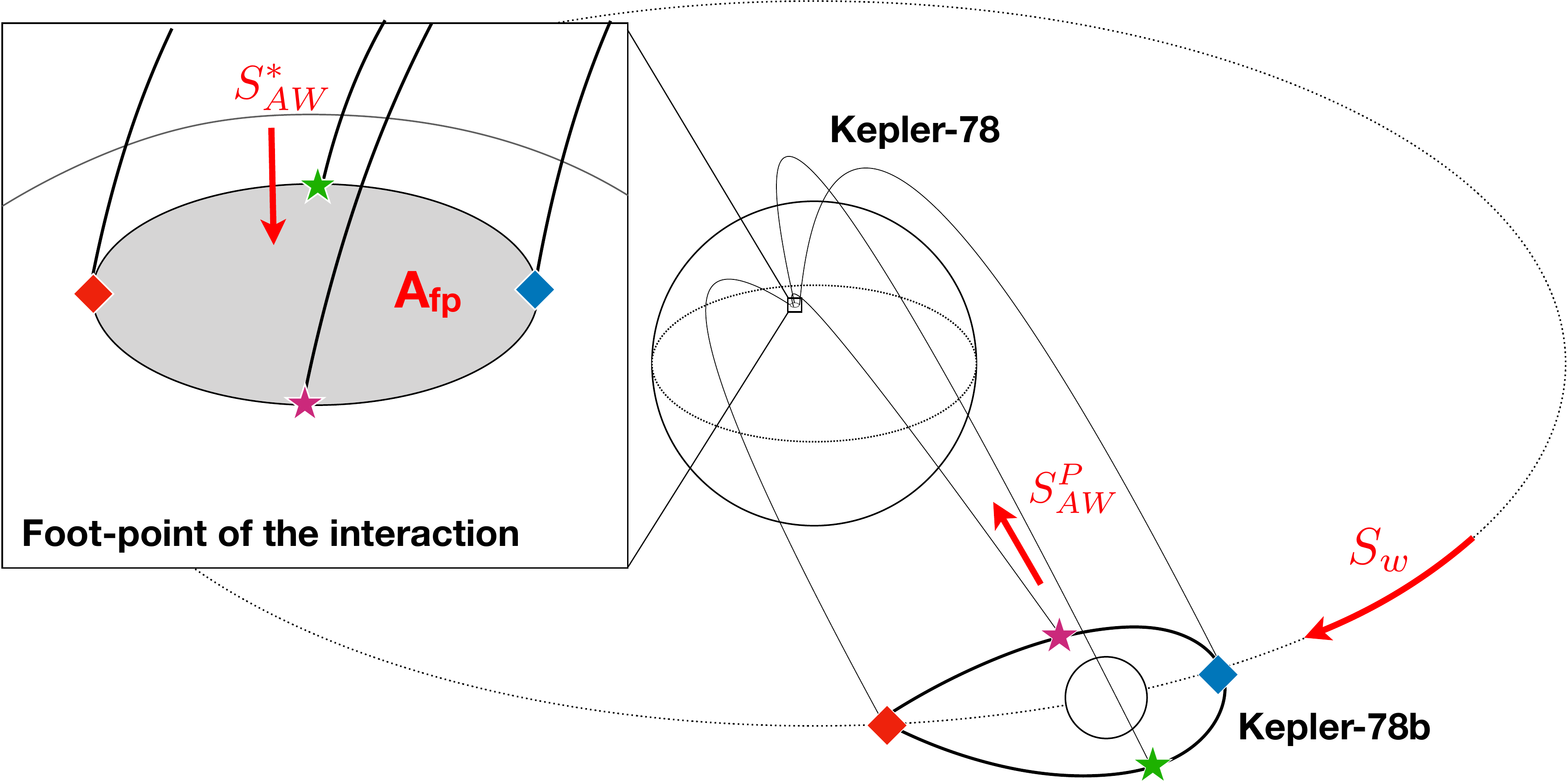}
  \caption{Schematic of the SPMI for a simplified magnetic topology. The shape of the planetary magnetosphere (bottom right) and of the area $A_{fp} $of the foot-point of the interaction (zoom on the left) are illustrative. Note that only one Alfv\'en wing is represented here (in the northern hemisphere), the second Alfv\'en wing also exists and impacts the southern hemisphere of the star.}
  \label{fig:SchematicFootPoint}
\end{figure*}

Due to the varying plasma conditions along the orbit, one may estimate
the lowest planetary magnetic field that is required for Kepler-78b to
sustain its own magnetosphere. The pressure equilibrium between the magnetosphere of the planet and its surrounding is characterized by the parameter 
\begin{equation}
    \Lambda_P=\frac{B_P^2}{2\mu_0P_t}\, ,
\end{equation}
where $B_P$ is
the magnetic field at the surface of the planet and $P_t$ the total
pressure in the stellar wind at the planetary orbit. A magnetosphere can be sustained only if $\Lambda_P \ge 1$. 
We display the planetary field estimate $B_P^{\rm min}$ such that $\Lambda_P=1$ in the last
panel of Figure \ref{fig:PropAlongOrbit}, for both the
3D wind and the starAML+PFSS model. \modd{This last panel actually also shows the wind magnetic field amplitude $B_w$ at the planetary orbit since we considered $\Lambda_P=1$ and the magnetic pressure dominates $P_t$ for planets on short-period orbit.} Overall, we predict that Kepler-78b needs a surface magnetic
field of at least 70 $\mu$T (0.7 G) to sustain a magnetosphere during a
complete orbit. For comparison, the Earth possesses an equatorial
magnetic field of about 30 $\mu$T (0.3 G) at its surface. 

The magnetic field of close-in planets is largely unknown as of
today. Theoretical works on planetary dynamos
\citep[\textit{e.g.} see the reviews of][]{Stevenson2003,Christensen2010} tend to show that the
magnetic field intensity primarily depends on the available power
inside the planet, and only marginally depends on the rotation rate. \moddd{This is particularly awaited in a saturated regime that can be expected for close-in planets such as Kepler-78b due to an efficient tidal locking.}
Kepler-78b is slightly larger and more massive than the Earth (see
Table \ref{ta:k78prop}, with large uncertainties on its mass). If its
internal structure is Earth-like, 
Kepler-78b is likely to possess a magnetic field slightly stronger
than the Earth and consequently a field that could be just strong
enough to maintain a magnetosphere. The available power for a dynamo
inside Kepler-78b can be furthermore much larger than in the Earth,
due to \textit{e.g.} more intense tidal interactions with its very close
host \citep[\moddd{note that this also depend on tidal-locking, for in-depth discussionss see}][]{Mathis2017b}. As a result we consider in the following that Kepler-78b sustains a strong enough magnetic field to possess a magnetosphere.

\subsection{Location of the interaction foot-point on the stellar
  surface}
\label{sec:locat-inter-footp}

The alfv\'enic perturbations due to SPMI follow the Alfv\'en characteristics
\reff{\begin{equation}
  \label{eq:alf_characteristics}
  {\bf c}_A^{\pm} = {\bf v}_o \pm {\bf v}_A\, 
\end{equation}}
and form the \textit{Alfv\'en wings} \citep{Neubauer1980}.
Following these characteristics from the planetary orbit using the 3D simulation of the stellar wind of Kepler-78, one can
trace back the impact location of the Alfv\'en wings on the stellar surface (see e.g. \citealt{Kopp2011}). This is shown schematically for the north-hemisphere Alfv\'en wing in Figure \ref{fig:SchematicFootPoint} where four Alfv\'en characteristics connecting the planet magnetosphere to the stellar surface are illustrated. The characteristic area $A_{fp}$ of the footpoint of the interaction is shown in the close-up of the stellar surface in the top-left insert. We will assume a characteristic size of the planet's magnetosphere based on the pressure equilibrium between the magnetic pressure due to the planetary magnetic field and the ambient pressure in the stellar wind. This approximation gives a satisfying estimate of the magnetospheric size at the front of the interaction in 3D simulations \citep{Strugarek2016c}, and generally under-estimate it in the magnetospheric tail. A more realistic estimate of the magnetospheric shape and size in the equatorial plane would not affect the main conclusions of our study.

\begin{figure*}[htb]
  \centering
  \includegraphics[width=\linewidth]{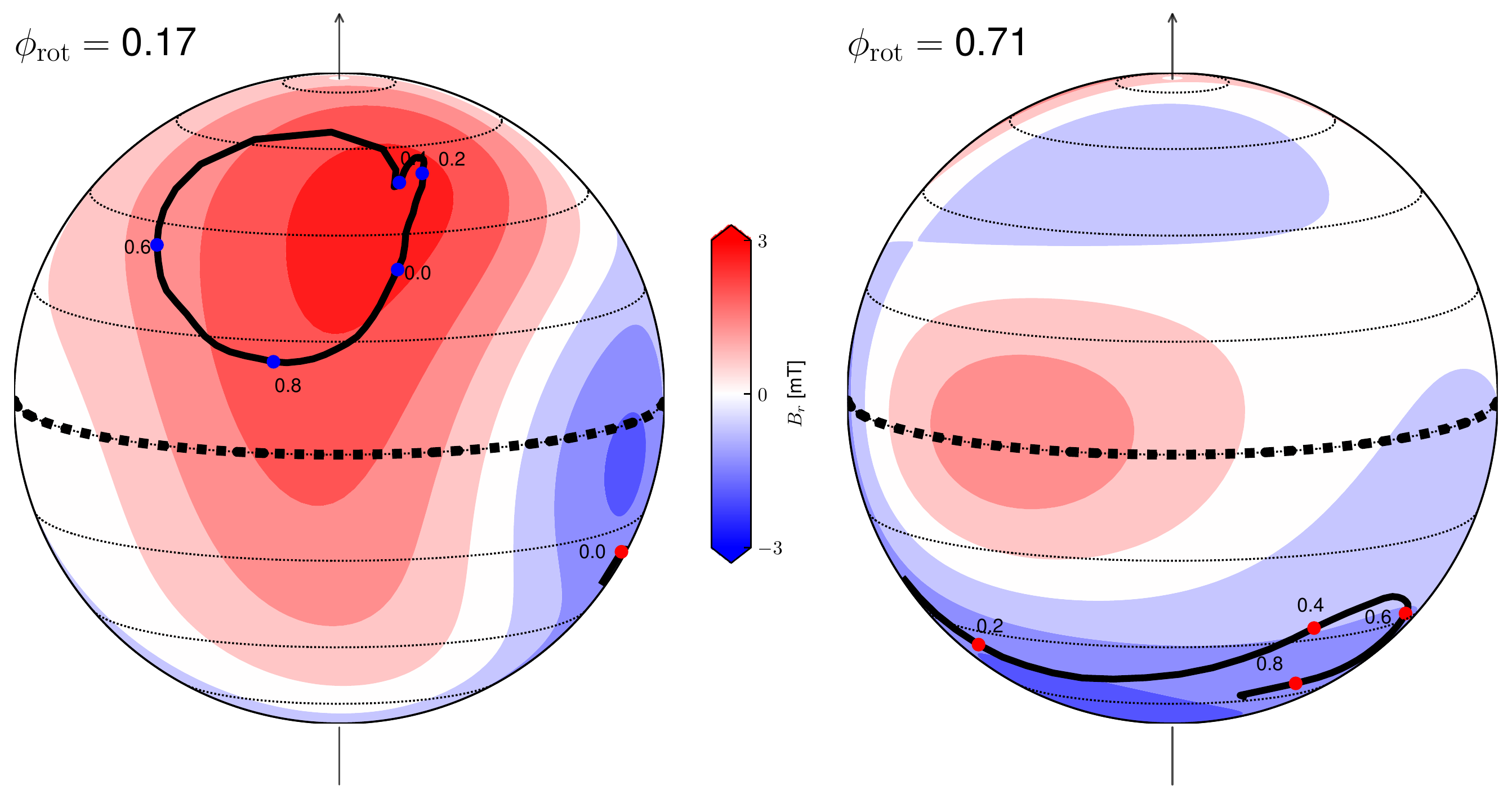}
  \caption{Foot-points of the magnetic interaction on the stellar surface seen for the two rotational phases of the star $\phi_{\rm rot} = 0.17$ (left) and $\phi_{\rm rot}=0.71$ (right). The radial magnetic field at the stellar surface is shown in red (positive) and blue (negative). The foot-point paths of the two Alfv\'en wings are represented by the black lines in each hemisphere. Five orbital phases are labeled along the path by the red and blue dots. Each dot thus represents the footpoint of the interaction (insert in Fig. \ref{fig:SchematicFootPoint}) at a particular orbital phase. \reff{The star is represented with a rotation axis (indicated by the black arrows) inclined by $i=80^\circ$ from the line-of-sight \citep{Moutou2016}. The equator of the star is labelled with a thick dashed line.}}
  \label{fig:FootPoints}
  \includegraphics[width=\linewidth]{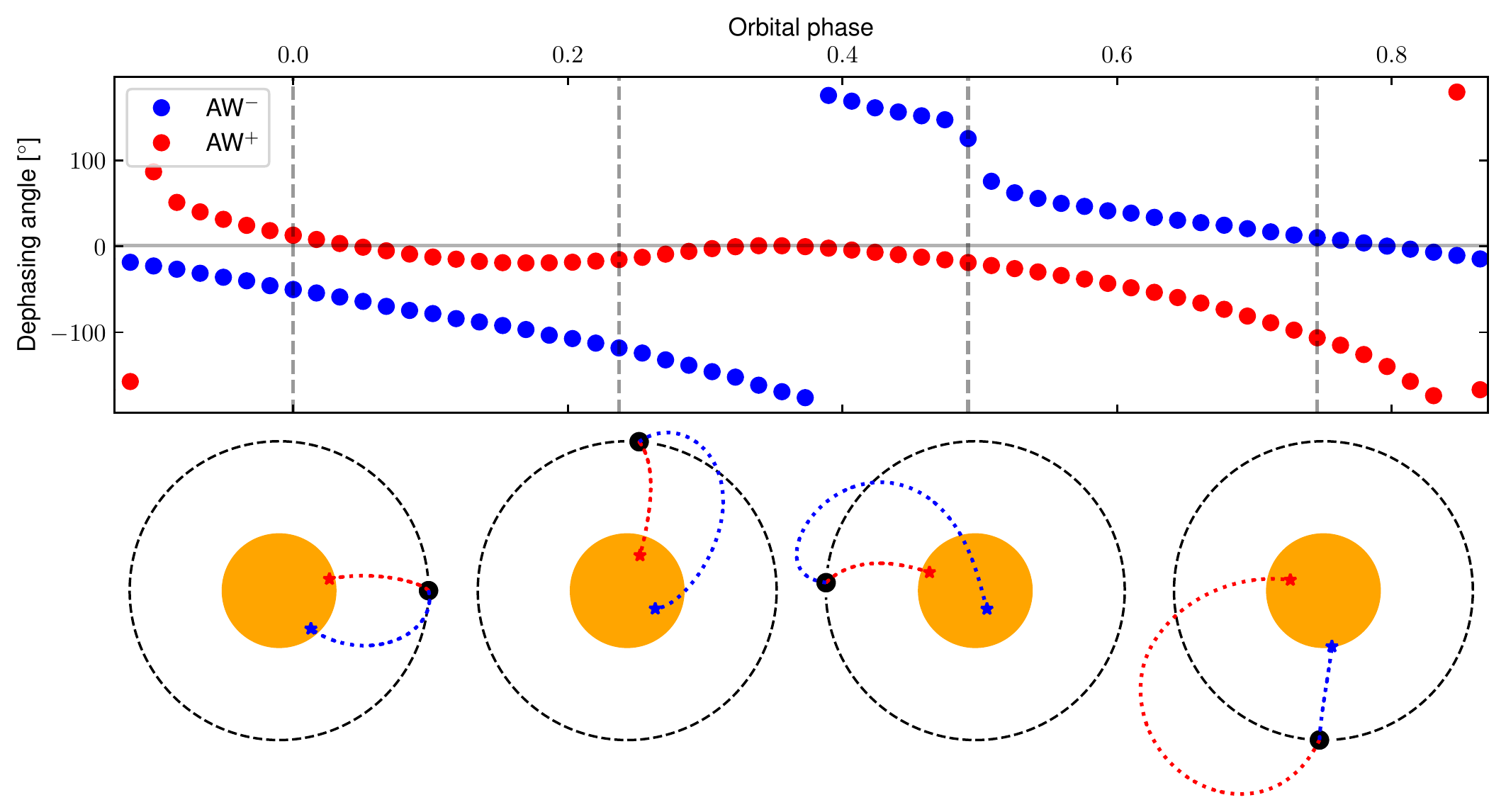}
  \caption{\textit{Top panel.} Dephasing angle between the orbital phase of the planet and the impact point of the Alfv\'en wings on the stellar surface, as a function of the orbital phase of the planet. \textit{Bottom panel.} Projection of the Alfv\'en wings on the orbital plane for four different orbital phases labelled by the dashed vertical gray lines in the top panel. In both panels, the northern Alfv\'en wing (AW$^-$) is shown in blue and the southern Alfv\'en wings (AW$^+$) in red.}
  \label{fig:phaselag}
\end{figure*}

We display the foot-points of the two wings (AW$^{-}$ and AW$^{+}$) in Figure
\ref{fig:FootPoints} for two different rotational phases of the star, along with the observed radial magnetic field
of the star at its surface in blue/red contours. As the planet orbits, the footpoints
of the wings move along the black lines on the stellar surface (the locations corresponding to the orbital phases 0., 0.2, 0.4, 0.6 and
0.8 are labeled by blue/red dots along path for the northern/southern Alfv\'en wing). We will refer to these paths as
the Foot-Point Paths (FPPs) later on. A front and a back view of
the stellar surface are shown in Figure \ref{fig:FootPoints},
which illustrates that each Alfv\'en wing primarily impacts opposite regions
of the stellar surface for this specific magnetic topology of Kepler-78 of July 2015. The star is viewed in
Figure \ref{fig:FootPoints} \moddd{with a rotation axis inclined at 80$^\circ$ to the line of sight}, which is the inclination angle of
Kepler-78 seen from the Earth. 

\modd{The characteristic area of the footpoints ($A_{fp}$) along the FPP depends on the
planetary field $B_P$ that sets the local size of magnetic interaction
in the planet vicinity (see Fig. \ref{fig:SchematicFootPoint}). For a planetary field of $100 \, \mu$T (1 G), the characteristic foot-point area is about $0.1\%$ of the disk of Kepler-78. Hence, the foot-point of the interaction extends on characteristic sizes comparable to standard sunspots.}

The magnetic topology of Kepler-78 and the subsequent location of the FPPs on the stellar surface has further important implications on the relative phase of the footpoint with respect to the planetary orbital phase. We illustrate this in Figure \ref{fig:phaselag} for the two Alfv\'en wings (top panel) as a function of the orbital phase, neglecting for the moment the rotation of the star. We remark that the phase lag between the planet and the footpoint of the interaction greatly varies along the orbit, especially for AW$^-$. This is of course a direct consequence of the compactness of the FPP on the stellar surface as shown on the left panel of Figure \ref{fig:FootPoints} for AW$^-$. The four schematics in the bottom row of Figure \ref{fig:phaselag} strikingly illustrate the phase-lag variations at four different phases along the orbit. \modd{We thus cannot expect the signal from SPMIs to be preferentially observed with a constant phase-lag with respect to the orbital phase and must rely on our global modelling to predict the detailed time-variability of SPMIs.}

\subsection{The combined effect of orbital motion and stellar rotation}
\label{sec:rot_and_orb}

As the star rotates, the FPP which is tied to the global magnetic
topology rotates as well. This simple observation has a
strong consequence on the tracers of SPMIs in the Kepler-78
system: any observable feature related to the SPMI in the spectrum of
Kepler-78 will thus be correlated to both the orbital period (displacement of the foot-point along the FPP) and the stellar rotation period that will set the visibility of the FPPs. How the two effects combine to produce an observable signal is explored in \S\ref{sec:mod-anomal-emission}. We already note that if the magnetic topology of
the star changes due to, \textit{e.g.}, a magnetic cycle, this feature
of the SPMI would also change accordingly. In addition, the
variations along the orbital path will lead to different
properties of the Alfv\'en wings along the FPP, which adds 
orbital-period modulations of any observable trace of the SPMI. 

\begin{figure}[htb]
  \centering
  \includegraphics[width=\linewidth]{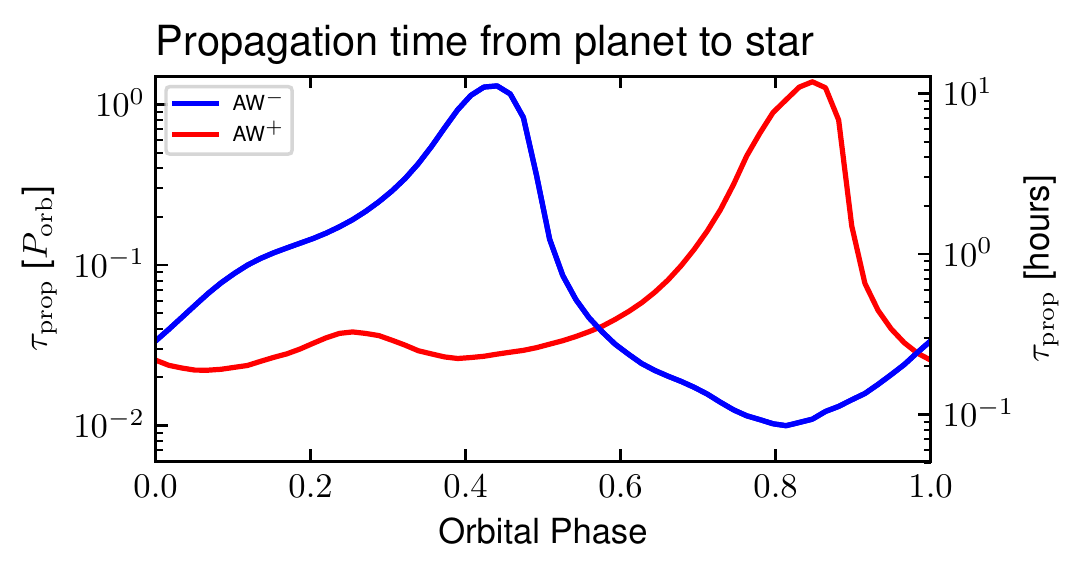}
  \caption{Propagation timescale between the orbit of the planet and the stellar surface as a function of the orbital phase. The timescale is normalized to the orbital period on the left axis and given in hours on the right axis. It is calculated separately for the two Alfv\'en wings (blue and red lines). A timescale of 1 on the left axis corresponds to a situation where the Alfv\'en waves reach the stellar surface by the time the planet has performed a full orbit.}
  \label{fig:PropTimeScale}
\end{figure}

An additional and often overlooked aspect further blurs the picture. 
The propagation timescale $\tau_{\rm prop}$ of Alfv\'en waves from the
planetary orbit down to the 
FPP depends on the plasma characteristics between the planet and
the star. As a result, perturbations triggered at different orbital phases can actually reach the stellar surface at the same time. To quantify this we integrate the inverse of the Alfv\'en speed along the
Alfv\'en characteristics ${\bf c}_A^{\pm}$ to estimate
$\tau_{\rm prop}$. The propagation timescale is shown in Figure \ref{fig:PropTimeScale} as a function of the orbital phase for
the two wings. During most of the orbit Alf\'ven waves reach the
stellar surface in a few percents of the orbital period, \textit{i.e.}
in about a few tens of minutes. At two noticeable orbital phases (around 0.4 for AW$^-$ and 0.8 for AW$^+$) the propagation time is much longer which leads to delays as large as the orbital period itself. 

The unipolar interaction model of \citet{Laine2012} requires that Alfv\'en waves can travel back and forth between the planet and the star, which could be the case here for a sub-part of the orbit. Conversely, the dipolar interaction model studied in
\citet{Strugarek2016c} does not require such a strong assumption \reff{and was robustly parametrized over a large range of star-planet configurations \citep{Strugarek2017e}}. Consequently we use
this latter model to estimate the energetic of the
star-planet magnetic interaction in the Kepler-78 system.

\section{Energetics of the star-planet magnetic interaction}
\label{sec:energ-star-plan}

\subsection{Magnetic power of the star-planet magnetic interaction along the orbit}
\label{sec:power-spmi}

We now estimate the energetics of the SPMI in the Kepler-78 system to
assess whether or not SPMIs could leave detectable signals in the
activity tracers of the star. We will base our analysis on the scaling
law for the Poynting flux in the Alfv\'en wings derived in
\citet{Strugarek2016c,Strugarek2017e} and which we recall here as
\begin{align}
  \mathcal{P} =& A_1\pi R_P^2 \cdot \left(1+C_1\cos\Theta_M\right) \cdot
  \left(\Lambda_P^\chi\right) \nonumber \\ \cdot & \left(c_d S_w M_a^{\xi_0\left(1+\xi_1\cos\Theta_M\right)}
    \right) \, ,
  \label{eq:PoyFlux}
\end{align}
where $A_1=2.5$, $C_1=0.83$, $\xi_0=0.44$, $\xi_1=-0.3$ and $\chi=0.25 $ are constants that were fitted with 3D numerical simulations in \citet{Strugarek2016c} and \citet{Strugarek2017e}, and $\Theta_M$ is the relative inclination of the planetary field with respect to the ambient magnetic field ${\bf B}_w$. The drag coefficient is defined as $c_d=M_a/\sqrt{1+M_a^2}$, and we left out any dependencies on the resistive properties of the planet ionosphere and stellar wind for the sake of simplicity (see \citealt{Strugarek2016c} for more details). \moddd{The approximate Poynting flux (\ref{eq:PoyFlux}) was shown to be accurate within 50\% when comparing the fully self-consistent 3D simulations \citep{Strugarek2016c} with analytical formulae derived by \citet{Saur2013}.}

We use Eq. \ref{eq:PoyFlux} to estimate the Poynting flux density in the planet vicinity
\begin{equation}
S_{AW}^{P} = \frac{\mathcal{P}}{\pi R_m^2}\, ,
\label{eq:SawP}
\end{equation}
where $R_m = R_P\Lambda_P^{1/6}$ is the \reff{size of the obstacle that consists in the planet and its magnetosphere. We also use Eq. \ref{eq:PoyFlux} to estimate} the Poynting flux impacting the stellar surface
\begin{equation}
    S_{AW}^{*} \simeq \frac{\mathcal{P}}{A_{fp}}\, .
    \label{eq:SawStar}
\end{equation}
We recall here that $A_{fp}$ is estimated by following the Alfv\'en characteristics from the magnetospheric borders in the vicinity of the planet down the stellar surface.

The three Poynting flux densities $S_w$, $S_{AW}^P$ and $S_{AW}^{*}$ are illustrated in Fig. \ref{fig:SchematicFootPoint}.
We suppose that the magnetic field of the planet is a dipole of strength $B_P$ at its equator, with a moment anti-aligned with the rotation axis. As a result, the inclination parameter in Eq. \ref{eq:PoyFlux} is simply $\Theta_M=\Theta_I$. This assumption makes the interaction in a $\sim 45^\circ$ inclination during most of the orbit. Note that a perfectly aligned configuration along the orbit would increase the Poynting flux amplitude (typically by a factor 2 at most, see \citealt{Strugarek2017}), but would not affect its temporal variability.

\begin{figure}[htb]
  \centering
  \includegraphics[width=0.95\linewidth]{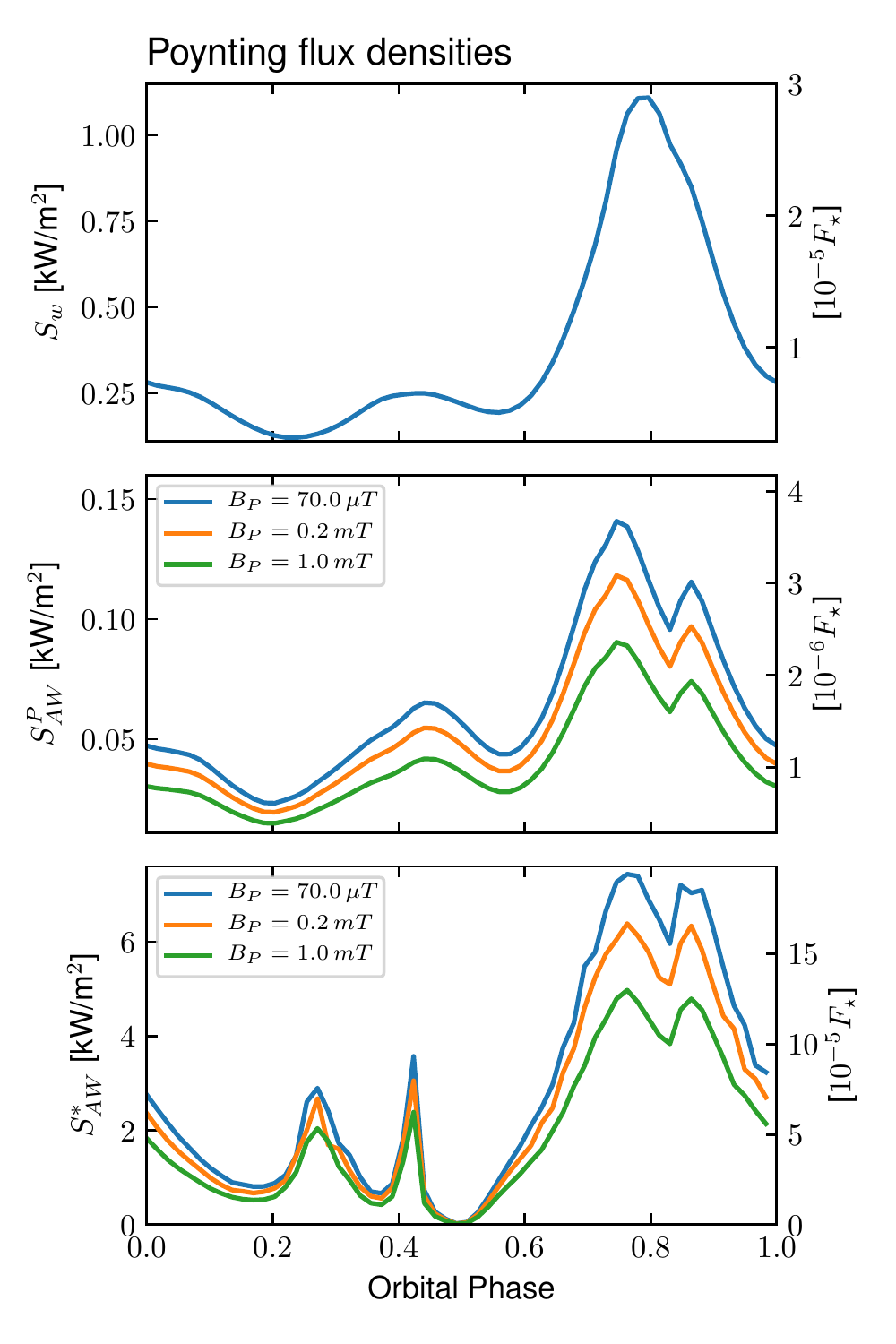}
  \caption{Poynting flux densities (see Fig. \ref{fig:SchematicFootPoint}) as a function of the orbital phase for three different planetary fields $B_P$. The fluxes are given in units of kW m$^{-2}$ on the left axis and in units of stellar bolometric flux on the right axis. The top panel shows the energy density in the stellar wind at the planetary orbit ($S_w$), the middle panel the Poynting flux density in the Alfv\'en wing in the planet vicinity ($S_{AW}^{P}$), and the bottom panel the Poynting flux density impacting the stellar surface ($S_{AW}^{*}$)}.
  \label{fig:PoyFlux}
\end{figure}

We display in Figure \ref{fig:PoyFlux} the three Poynting flux densities $S_w$, $S_{AW}^P$ and $S_{AW}^{*}$ as a function of the orbital phase for three different planetary field
intensities. The planet-wind interaction gives a 10\% conversion efficiency from the incoming Poynting flux $S_w$ (top panel) to the Poynting flux in the vicinity of the planet $S_{AW}^P$ (middle panel). In spite of its small size, the magnetic interaction of Kepler-78b leads to Poynting flux densities $S^P_{AW}$ of the order of $10^{-2}-10^{-1}$ kW m$^{-2}$ for realistic magnetic field intensities. The Poynting flux density weakly decreases with $B_P$ ($S_{AW}^P \propto B_P^{-0.1}$) due to the competition of two opposite effects. As $B_P$ increases, the magnetospheric size increases, allowing a larger area of interaction that increases $\mathcal{P}$. At the same time, increasing the area of interaction decreases the flux density. The second effect is slightly stronger which leads to the observed weak negative dependency with $B_P$. Note that this dependency is sensitive to a correct estimation of the non-spherical magnetospheric size of Kepler-78b, which is beyond the scope of the present work. 

As the alfv{\'e}nic perturbations follow converging magnetic field lines when propagating towards the star, the Poynting flux density increases up to a few kW m$^{-2}$ when it impacts the stellar lower atmosphere ($S_{AW}^{*}$, bottom panel of Fig. \ref{fig:PoyFlux}). \refff{The observed H$\alpha$ flux at $\phi_{\rm rot}=0$ (see Fig. \ref{fig:equivlums}) corresponds to a power density of 60 kW/m$^2$ if it originates from a localized area of 0.1\% of the stellar disk (averaged footpoint area, see above). As a result, the predicted SPMI flux density is one order of magnitude smaller than the observed flux density. In addition, it must be noted that only part of the Alfv\'en wings energy will likely be converted to the activity band. For instance, in accretion-powered emissions in young systems the conversion factor in H$\alpha$ is typically of a few percents \citep[e.g.][]{Fang2009}. If such a conversion factor applies to SPMIs, their detection would be challenging but could be foreseen provided a distinctive temporal modulation can be identified. We now describe how to predict the temporal variability of SPMI signals with the example of Kepler-78.} 


\begin{figure*}[htb]
  \centering
  \includegraphics[width=\linewidth]{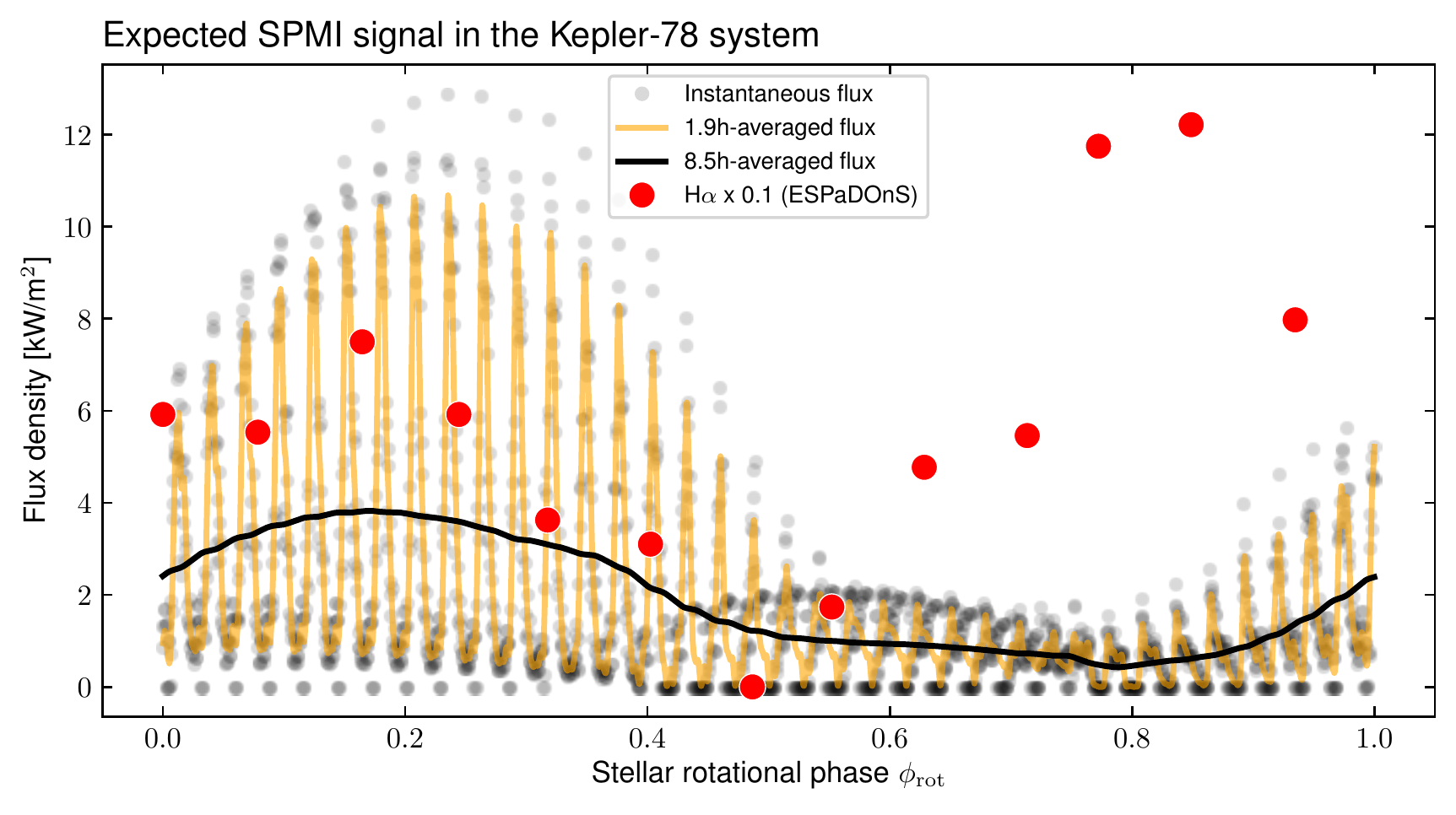}
  \caption{Instantaneous flux density (transparent gray dots) as a function of the rotational phase of the star. We also show the SPMI signal smoothed over 1.9 hours (orange) and over the orbital period (black). \refff{The observed H$\alpha$ flux is shown by the red circles. It is the same data as in Fig. \ref{fig:equivlums}, multiplied by 1000 to mimic a signal originating from the footpoints of the SPMI (covering on average 0.1\% of the stellar disk), and then divided by 10 to easily compare the temporal modulations of the observed and predicted signals.}}
  \label{fig:EmissionsVariations}
\end{figure*}

\subsection{Modeling the SPMI emission modulation}
\label{sec:mod-anomal-emission}

\modd{The power channeled in the SPMI can in principle be transferred into an observable signal through various mechanisms \citep[e.g. see][and references therein]{Saur2017}. Nevertheless, the timing of any SPMI-induced feature is at first order independent of the detailed physical mechanism producing it: it is determined by the topology of the Alfv\'en wings (Fig. \ref{fig:FootPoints}) and the Alfv\'en waves travel time (Fig. \ref{fig:PropTimeScale}), which can be predicted from the present simulations.} Indeed, the foot-point of the interaction sweeps over the stellar surface as the planet orbits, as shown in Fig. \ref{fig:FootPoints}. In the meanwhile the star itself rotates as well, which induces a complex temporal variability.

The exact mechanism leading to an observable tracer nonetheless matters: emissions from accelerated particles will likely be isotropic, while local heating of the chromosphere/photosphere will likely be subject to limb-darkening effects. In the latter case, we need to correct the expected SPMI flux along the FPP with limb-darkening based on the quadratic law of \citet{Espinoza2015} (the coefficients for Kepler-78 were obtained using the \href{https://github.com/nespinoza/limb-darkening}{limb-darkening python-package} by \citealt{Espinoza2015}, with closest match being $T_{\rm eff} = 5000 $ K, log g = 4.5, [Fe/H] =0 and micro-turbulent velocity of 2 km/s). We have included the effect of limb-darkening in our modelling in what follows. Nevertheless, our conclusions hold with and without limb-darkening effects.

The resulting expected \modd{instantaneous} emission variability in the H$\alpha$ band is shown in Fig. \ref{fig:EmissionsVariations} by gray dots, along with the \modd{1.9 hours-averaged (orange, corresponding to the observational exposure time, see \citealt{Moutou2016}) and orbit-averaged (black) emissions}. We assumed here a planetary field $B_p=50\mu T$ (blue curves in Fig. \ref{fig:PoyFlux}) and computed the emission for one rotational phase of the star, i.e. about 34 orbits of Kepler-78b. The \reff{fast variations in the gray dots and in the orange curve mainly originate from the planetary orbital motion through the inhomogeneous corona of the star.} The orbit-averaged signal (black curve) naturally filters the orbital motions and shows a clear modulation with the rotational period of the star, as expected. 

\refff{We overlay the observed flux in the H$\alpha$ band (red circles, see also Figure \ref{fig:equivlums}), multiplied by a factor of 1000 to mimic an emission originating from the SPMI footpoints (over an average area of 0.1\% of the stellar surface), and divided by 10 to ease the comparison between the two signals. We remark that the temporal modulation of the observed signal is not well correlated with the SPMI signal. In particular the excess flux around $\phi_{\rm rot} = 0.8$ corresponds to the minimum excess flux awaited from the SPMI.}

\refff{Interestingly, a dephasing of $\sim$0.3 in the rotational phase of the star is enough to restore an excellent correlation between the observed signal and the SPMI flux modulation. The magnetic field topology of Kepler-78 determines the temporal modulation of the SPMI flux. As a result, dephasing the SPMI signal by 0.3 would require major changes in the magnetic field topology (see Appendix \ref{sec:param-sensitivity}).} \modd{We have assessed the robustness of the variability of the SPMI signal along the orbital phase (black line) by varying the inclination angle of Kepler-78 with respect to the Earth, the orbital period of Kepler-78b, and the number of spherical harmonics used in the ZDI magnetic map. The detailed results are shown in Appendix \ref{sec:param-sensitivity}. The main conclusion is that the overall trend with the rotational phase is extremely robust.}

\modd{Finally, we note that the three last points of observation were taken about 2.5 rotation periods later during the summer of 2015. It is possible that the magnetic topology of Kepler-78 changed significantly over this period of time, with for example the appearance/disappearance of active regions affecting the large-scale magnetic topology \reff{or due to changes through the shear of the differential rotation}. A detection of SPMI would thus require denser observational datasets to properly address this uncertainty.}

\section{Discussions on the detectability of star-planet magnetic interactions}
\label{sec:discussions}

\modd{We have shown that SPMI in the Kepler-78 system is powerful enough to be \refff{potentially} detectable in characteristic activity tracers such as H$\alpha$ \citep{Moutou2016}. 
Our study opens up promising possibilities of detection for more favorable systems and/or through dedicated and carefully planned observational campaigns. We discuss here these aspects to pave the road towards future positive detections of SPMIs.} 

\textit{Stellar magnetic field.} First and foremost, we reiterate that several critical ingredients are needed to properly hunt for traces of star-planet magnetic interactions: (i) the magnetic topology of the star, (ii) the mass and radius of the host star, and (iii) the orbital properties of the planet. Once these three aspects are properly constrained, one can relate a SPMI emission to the magnetic properties of the orbiting planet. Without these three aspects, it is impossible to predict how any signal for the SPMI is modulated as a function of, \textit{e.g.}, the orbital period of the planet or the rotational period of the star. 
It is further useful to recall that the SPMI emission increases with the stellar magnetic field. Indeed, extracting to first order the dependency in $B_w$ in Eq. \ref{eq:SawP}, one finds $S_{AW}^P \propto B_w^{0.7}$. More favorable systems with large stellar magnetic field can thus be expected to show strong SPMI signals. 



\textit{Planetary magnetic field.} The amplitude of the SPMI emissions also depends on the assumed planetary field amplitude and topology. \modd{Note that at first order, the temporal variability of the signal itself does not depend on the planetary field.} The Poynting flux $\mathcal{P}$ channeled by the magnetic interaction is roughly proportional to $B_P^{1/2}$ (see Eq. \ref{eq:PoyFlux}). As $B_P$ increases, the size of the magnetosphere increases and the shape and strength of the Alfv\'en wings are slightly altered (see Fig. \ref{fig:PoyFlux}), leading to small changes in the properties of the foot-point of the interaction. We expect giant close-in planets with strong magnetic field to be the most favorable targets to detect SPMI tracers. 

\textit{Stellar wind model uncertainties.} The analysis carried in this work relies on the wind modelling of the central star. Parameters such as the coronal temperature and density, or equivalently the mass loss rate of the star, are not well constrained for Kepler-78. We have assessed the sensitivity of our wind modelling choice by varying the base density of the model ($n$, see Eq. \ref{eq:rho_and_T}) by a factor of 4. This affects the alfv\'enic Mach number in the stellar wind and the amplitude of the SPMI flux density can thus vary by the same factor. Nevertheless, the temporal variability of the SPMI signal is found to be very robust with respect to the plasma properties at the base of the corona (see also Appendix \ref{sec:param-sensitivity}). 

\modd{\textit{Observational strategy.} We have shown that any signal of SPMIs in the Kepler-78 system should be modulated over both the orbital and the stellar rotational periods. The unambiguous detection of SPMI thus requires denser observational campaigns covering both the orbital timescale (in-and-out of transit observations) and the rotational timescale (recurrent visits over a timescale of weeks). When the magnetic topology of the star is known not to vary much over a given observational period, our methodology allows us to predict the favorable rotational and orbital phases for SPMI signals. These interesting phases can be predicted combining simple stellar wind models with potential-field source-surface extrapolations (e.g. with our simplified but calibrated starAML+PFSS package, see \S\ref{sec:simul-magn-kepl}), i.e. without relying on computationally-expensive modelling. As a result, observational campaigns to detect SPMIs can leverage the known properties of the star-planet systems to identify on-the-fly the preferred observation windows where the signal is the most likely to be detected.}


The analysis presented in this paper opens a promising avenue to estimate the magnetic properties of an exoplanet, provided (i) we have the necessary observational constraints on the system and (ii) a signal well-correlated with the expected time variability of the SPMI can be unambiguously detected. Favorable star-planets system would consist in a central dwarf or young star with a strong, dipolar-like magnetic field (\textit{e.g.} \`a la AD Leo, see \citealt{Morin2008}) around which a very close-in hot Jupiter would orbit. We intend to analyze more systematically such systems in the near future. 

\section{Conclusions}
\label{sec:conclusions-1}

We have applied in this work the concept of Alfv\'en wings to star-planet magnetic interactions in the Kepler-78 star-planet system. We modelled the environment of Kepler-78 with a 3D wind model \citep{Reville2016a} based on the observed magnetic topology of the star with Zeeman-Doppler-Imaging \citep{Moutou2016}. We have shown that the environment of Kepler-78b varies along its orbit which leads to large modulations of the magnetic interaction. Furthermore these modulations lead to time-lags between observable SPMI features along one orbit of Kepler-78b. Consequently, signals from SPMI  have to rely on a dedicated modelling of the star-planet system leveraging the orbital properties of the planet as well as the magnetic properties of the star.

Despite the relative small size of Kepler-78b, we have shown that SPMI can channel energy flux densities up to a few kW $m^{-2}$ toward Kepler-78. This flux is a priori large enough to be \refff{detectable with present telescope capabilities in Kepler-78 (see Fig. \ref{fig:EmissionsVariations} and \citealt{Moutou2016}), provided enough energy is transferred from the SPMI to the activity tracer considered.}

The detailed modelling of the wind of Kepler-78b allows us to go beyond order-of-magnitude estimates and predict the temporal variability awaited from SPMIs (see Fig \ref{fig:EmissionsVariations}). \refff{We show that the most favorable period for detecting occurs around $\phi_{\rm rot}=0.25$ for the magnetic topology of Kepler-78 as of July 2015. We have shown that the temporal variability predicted by our model is very robust to typical uncertainties in the planet orbital characteristics, inclination angle, as well as ZDI map spatial resolution (see Appendix \,\ref{sec:param-sensitivity}).} 

We have focused here our analysis on reproducing the temporal variability of standard magnetic activity tracers of a star. It is nevertheless not obvious today which wave-bands should be the most favorable to detect SPMIs. Our modelling allows us to characterize the available power in SPMIs and its temporal signature. How is this power transmitted from Alfv\'en waves to radiation? These waves can transfer their energy by wave-wave or wave-turbulence interactions in the low corona of the star; can accelerate energetic particles as in the magnetosphere of Jupiter; or can even act as a trigger of stellar flares. A careful scrutiny of these different mechanisms and their subsequent preferential observable wavebands (radio, X, visible, UV...) is today needed to further probe SPMIs in distant exoplanetary systems.

The methodology presented in this work nevertheless opens a clear path for the future detection of SPMIs. Denser observational campaigns of star-planet systems covering targeted orbital and rotational phases would allow to unambiguously separate stellar activity from SPMIs in favorable star-planet systems such as Kepler-78. The chase for star-planet magnetic interactions is still at its early stages and we can foresee their systematic detection in the coming years.

\acknowledgments

\reff{The authors thank an anonymous referee for valuable comments that improved the analysis presented in this paper.}
A.S. thanks A. Mignone and his team for giving the PLUTO code to the research community and J. Saur for discussions about star-planet and planet-satellite magnetic interactions. A.S. acknowledges access to supercomputers through GENCI (grant 20410133). A.S. and A.S.B. acknowledge funding from CNES-PLATO and CNES-Space Weather grants. A.S. and A.S.B. acknowledge funding from the Programme National Soleil-Terre (PNST). A.S. acknowledges funding from the Programme National de Plan\'etologie (PNP). J.F.D. acknowledges funding from the European Research Council (ERC) under the H2020 research \& innovation program (grant agreement \#740651 NewWorlds).

\bibliographystyle{yahapj}

\appendix

\section{Sensitivity of the modelled SPMI signal}
\label{sec:param-sensitivity}

\begin{figure*}[htb]
  \centering
  \includegraphics[width=0.49\linewidth]{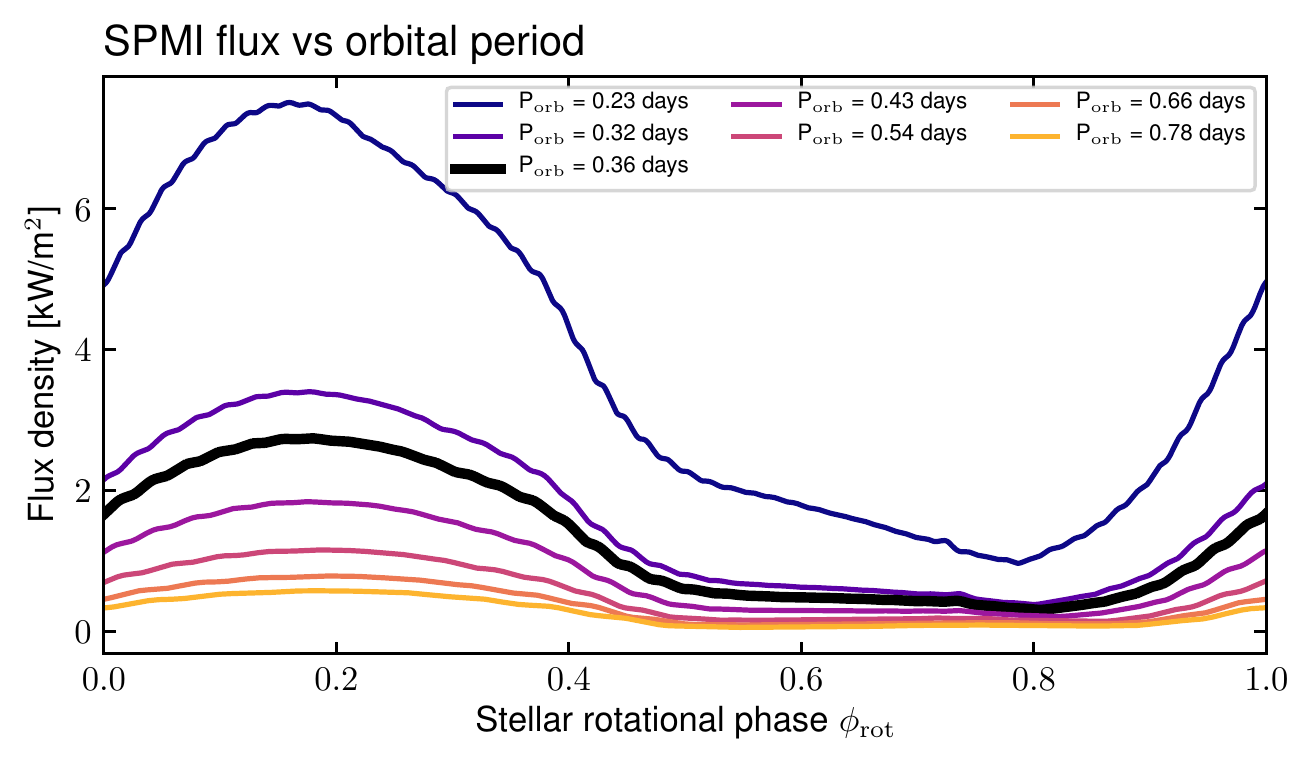}
  \includegraphics[width=0.49\linewidth]{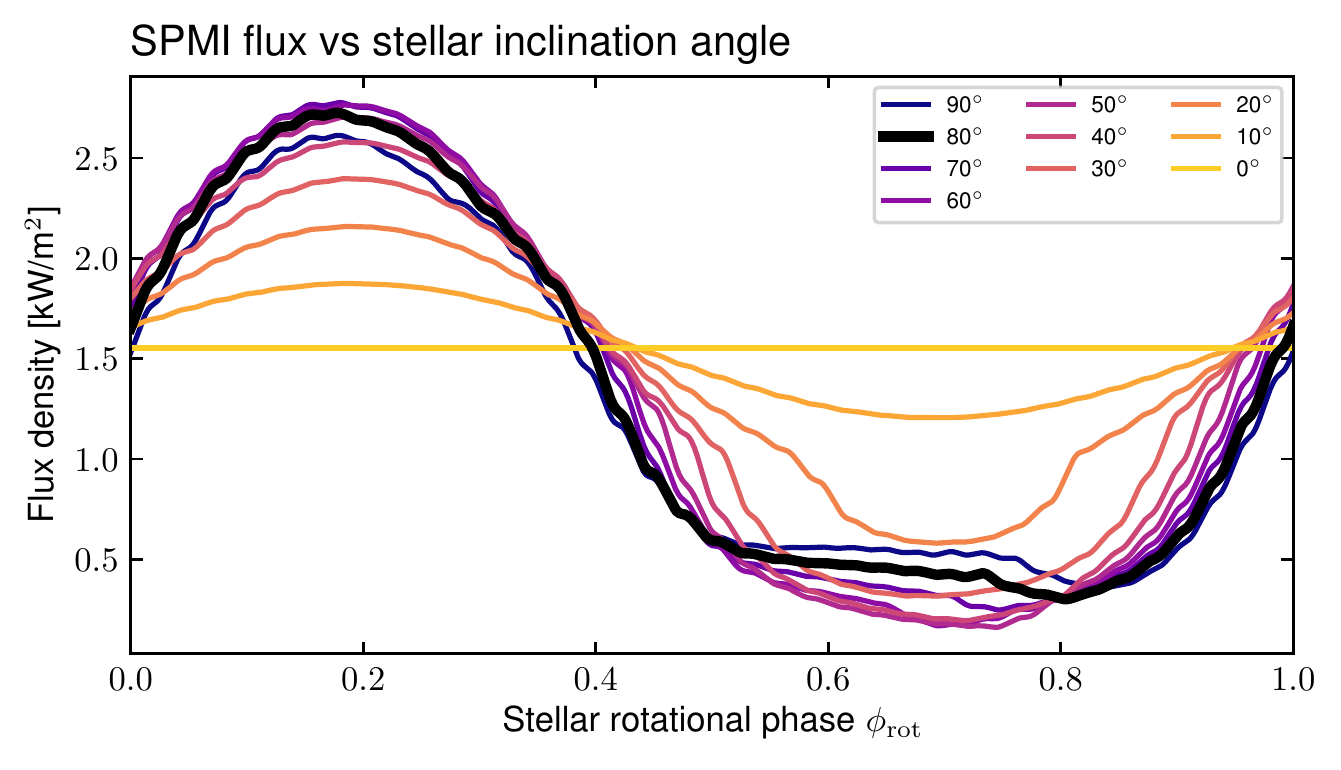}
  \includegraphics[width=0.49\linewidth]{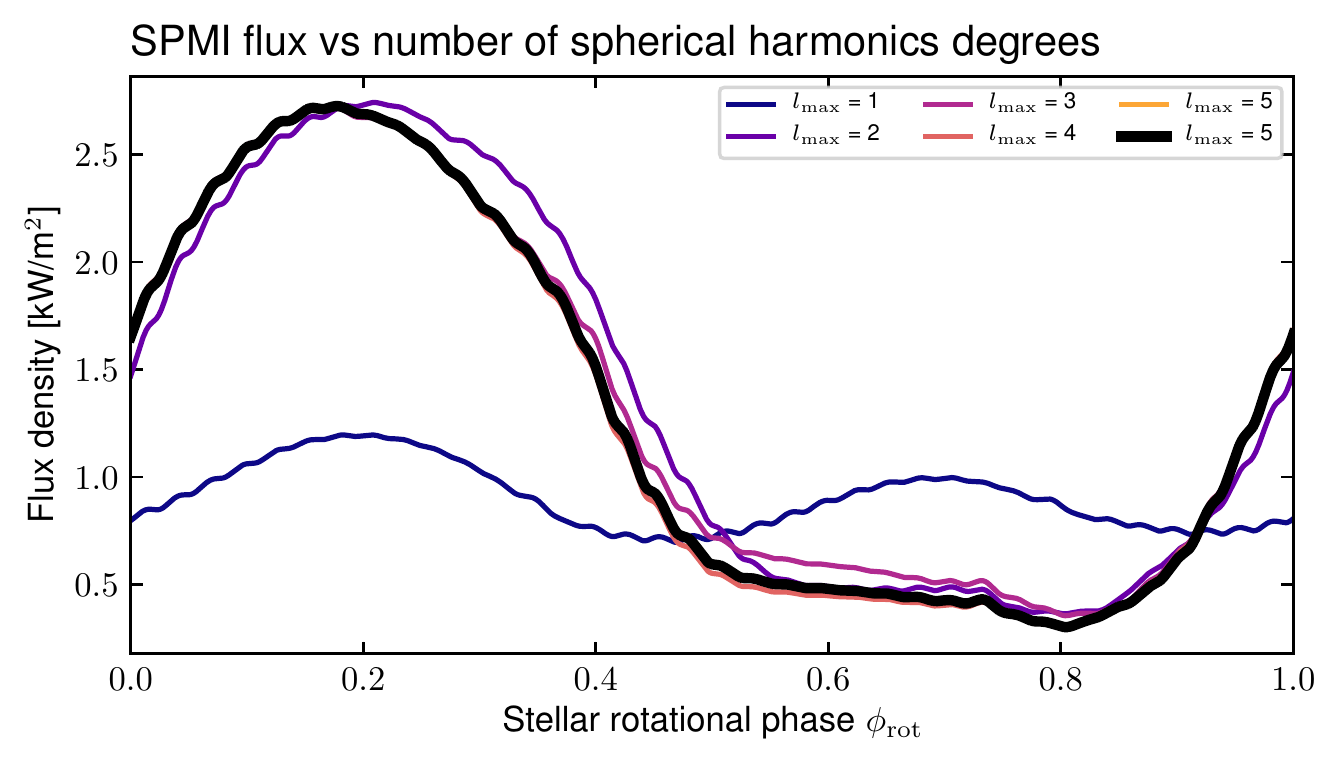}
  \caption{Parameter sensitivity study of the modelled SPMI signal. In each panel, the orbit-averaged awaited SPMI signal is shown as a function of the stellar rotation phase. The reference model studied in this paper, \textit{i.e.} the black line in Fig. \ref{fig:EmissionsVariations}, is also shown as a black line in the different panels. \textit{Top left.} Predicted fluxes for orbital periods varying from 0.23 to 0.78 days, which corresponds to semi-major axes from 2. to 4.5 $R_\star$. \textit{Top right.} Predicted flux for inclination angles of Kepler-78 (with respect to the line-of-sight) from 0$^\circ$ to 90$^\circ$. \textit{Bottom.} Predicted flux for a magnetic map of Kepler-78 reconstructed with spherical harmonics degrees $l$ up to $l_{\rm max}$, with $l_{\rm max} \in [1,5]$. }
  \label{fig:Sensitivity}
\end{figure*}

\modd{The modelling of SPMIs in Kepler-78 is subject to several observational uncertainties. We explore the sensitivity of our results in this Appendix and show that the overall temporal modulation of the awaited SPMI signal is very robust for this star-planet system if we neglect strong temporal variability of the stellar magnetic field.

We have use our calibrated starAML+PFSS model to explore the sensitivity of our results. We have tested three uncertain observational parameters: the orbital period of the planet, the inclination angle of the rotation axis of Kepler-78 with respect to the Earth, and the number of spherical harmonics used in the ZDI magnetic map. For each of these parameters, we applied our analysis pipeline and compare the orbit-averaged signal (black line in Fig. \ref{fig:EmissionsVariations} and in the panels of Fig. \ref{fig:Sensitivity}) due to SPMI. 

We have varied the orbital period of Kepler-78b from 0.23 days to 0.78 days (upper left panel of Fig. \ref{fig:Sensitivity}) which corresponds to semi-major axes from 2 to 4.5 stellar radii. The amplitude of the SPMI varies by an order of magnitude. This simply reflects the change in the amplitude of the stellar wind magnetic field amplitude at the planetary orbit that sets the Poynting flux $S_w$ at the source of the magnetic interaction. Nonetheless, the temporal modulation of the signal is very robust, with a maximum around $\phi_{\rm rot}\simeq 0.15$  and a minimum around $\phi_{\rm rot}\simeq 0.75$ for all orbital periods. \reff{Note that this exploration is wider than the error-bars on the orbital period of Kepler-78b \citep{Sanchis-Ojeda2013} in order to demonstrate the robustness of the temporal modulation of the SPMI signal.}

Next we have changed the inclination angle of the rotation axis of Kepler-78 with respect to us from $i=0^\circ$ to $i=90^\circ$ (upper right panel in Fig. \ref{fig:Sensitivity}). Note that this exploration is far wider than the error bars on the inclination angle and is intended to be educational. This time the effect is mainly on the temporal modulation itself. In the extreme case of a perpendicular inclination $i=0^\circ$, we always see the same pole of the star. As a result, we always see the same Alfv\'en wing foot-point path and the orbit-averaged SPMI signal does not show any temporal modulation, as expected. Nonetheless, the temporal modulation still shows the same robust trend as the inclination angle increases, albeit with a smaller amplitude.

Finally, we also have explored the impact of the resolution of the magnetic map on the estimated SPMI signal. We have carried out the analysis by limiting the number of spherical harmonics degrees $l$ to $l_{\rm max}$, with $l_{\rm max} \in [1,5]$ ($l_{\rm max}=5$ corresponds to the magnetic map used in this paper). The result is shown in the lower left panel of Fig. \ref{fig:Sensitivity}. We remark that the predicted flux does not change significantly when $l_{\rm max}\ge 2$. 
We recall that the last three observational points (for late rotational phases) were taken at the end of August 2015, a month after the other observations. We thus conclude that significant changes of the magnetic topology of Kepler-78 (in particular the $l=2$ components) over this period of time is required to accommodate these late observational points with the SPMI scenario.}

\section{Activity tracers for Kepler-78}
\label{sec:activ-trac-kepl}

\begin{figure}[htb]
  \centering
  \includegraphics[width=0.49\linewidth]{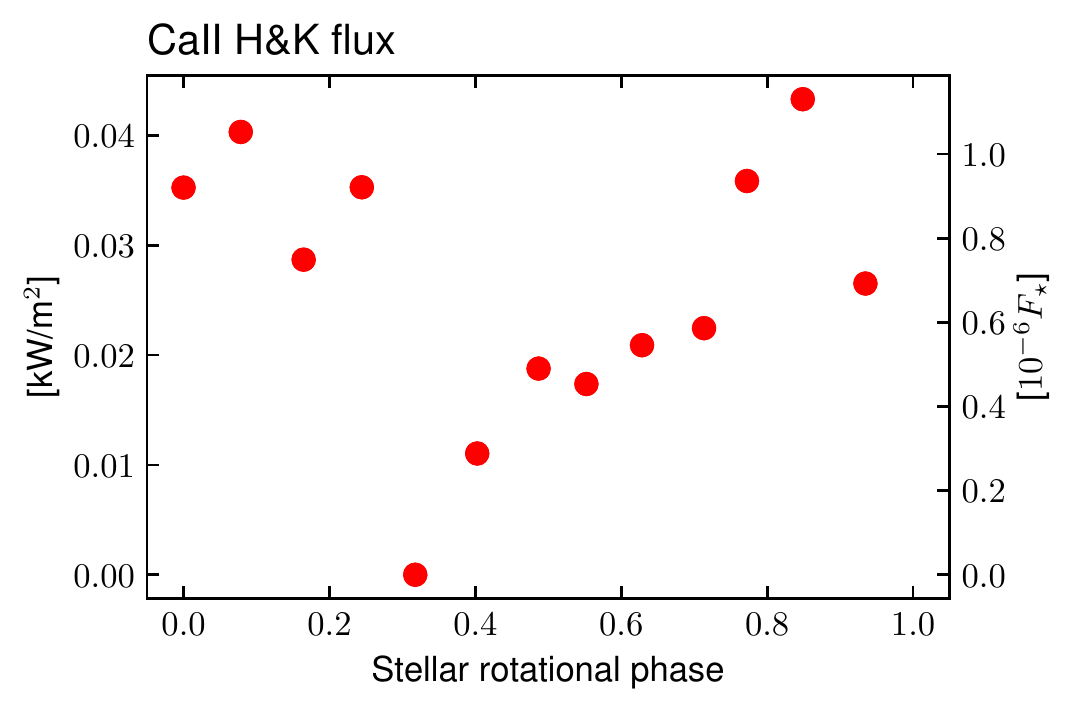}
  \includegraphics[width=0.49\linewidth]{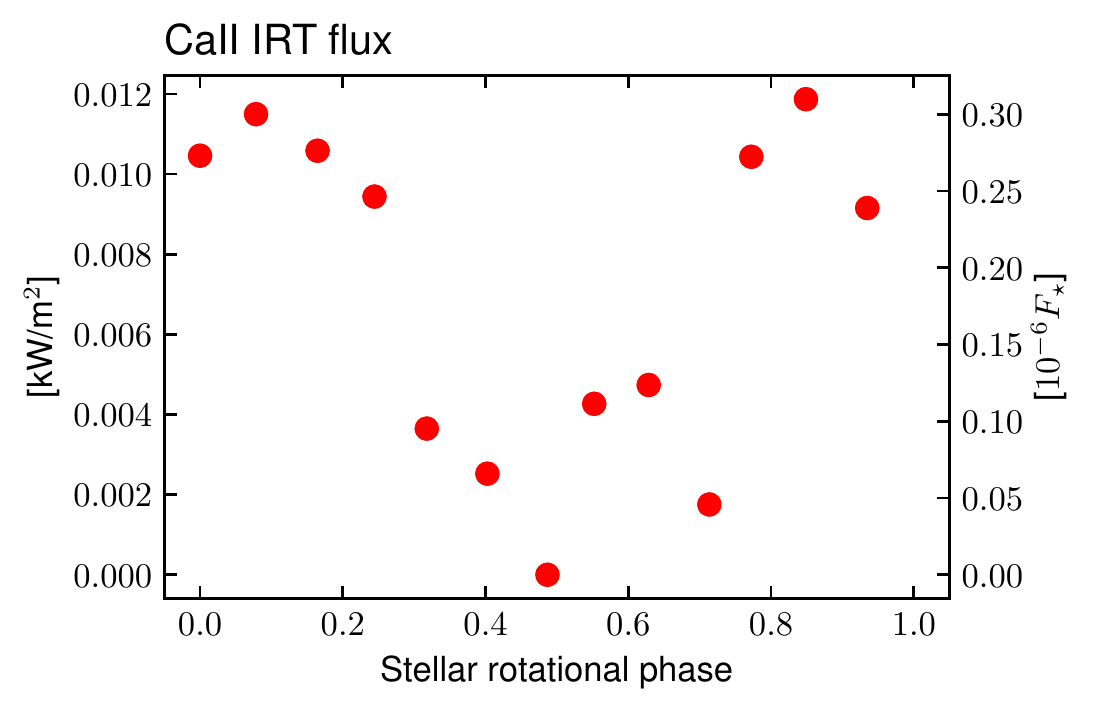}
  \caption{\reff{Flux observed in the H\&K bands (left) and infra-red triplet (right) of CaII as reported by \citet{Moutou2016}. The layout of the figures is the same as in Fig. \ref{fig:equivlums}.}}
  \label{fig:CaII}
\end{figure}

\reff{We have focused our study on one particular activity tracer (H$\alpha$) of Kepler-78. In the original work of \citet{Moutou2016}, several other activity tracers including the H\&K bands and infra-red triplet (IRT) of CaII were also characterized. We report in Fig. \ref{fig:CaII} these tracers in the same layout as in Fig. \ref{fig:equivlums}. We remark that the overall rotational modulation of these tracers is similar to H$\alpha$. The flux in the H\&K bands is of similar amplitude than H$\alpha$, and the IRT flux is about four to five times smaller. The main difference lies in the rotational phase where the observed flux is minimized: it occurs around $\phi_{\rm rot} = 0.5$ for H$\alpha$, $\phi_{\rm rot}=0.3$ for CaII H\&K, and $\phi_{\rm rot}= 0.5$ for CaII IRT. In all bands, the maxima always occur near $\phi_{\rm rot} \sim 0.15$ and $\phi_{\rm rot} \sim 0.8$. As a result, we see that the main conclusions of this work are not affected by the differences between the observed flux in the various activity tracers characterized by \citet{Moutou2016}.}

\end{document}